\documentclass[12pt]{article}%
\usepackage{amsfonts}
\usepackage{amsmath}
\usepackage{amssymb}
\usepackage{graphicx}%
\providecommand{\U}[1]{\protect\rule{.1in}{.1in}}
\newtheorem{theorem}{Theorem}[section]

\newtheorem{definition}[theorem]{Definition}
\newtheorem{example}[theorem]{Example}

\newtheorem{lemma}[theorem]{Lemma}
\newtheorem{notation}[theorem]{Notation}

\newtheorem{proposition}[theorem]{Proposition}
\newtheorem{remark}[theorem]{Remark}

\newtheorem{Examples}[theorem]{Examples}
\newenvironment{proof}[1][Proof]{\noindent\textbf{#1.} }{\ \rule{0.5em}{0.5em}}
\begin{document}

\title{The Measure of a Measurement}
\author{Palle E. T. Jorgensen\\The University of Iowa}
\date{}
\maketitle

\begin{abstract}
We identify a fractal scale $s$ in a family of Borel probability measures
$\mu$ on the unit interval which arises idependently in quantum information
theory and in wavelet analysis. The scales $s$ we find satisfy $s\in
\mathbb{R}_{+}$ and $s\not =1$, some $s\,<1$ and some $s>1$. We identify these
scales $s$ by considering the asymptotic properties of $u\left(  J\right)
/\left\vert J\right\vert ^{s}$ where $J$ are dyadic subintervals, and
$\left\vert J\right\vert \rightarrow0$.

\end{abstract}

\section{Introduction}

While finite non-commutative operator systems lie at the foundation of quantum
measurement, they are also tools for understanding geometric iterations as
used in the theory of iterated function systems (IFSs) and in wavelet
analysis. Key is a certain splitting of the total Hilbert space and its
recursive iterations to further iterated subdivisions. This paper explores
some implications for associated probability measures (in the classical sense
of measure theory), specifically their fractal components.

In quantum communication (the study of (quantum) error-correction codes),
certain algebras of operators and completely positive mappings form the
starting point; see especially the papers \cite{KLPL06} and \cite{Kri05}. They
take the form of a finite number of channels of Hilbert space operators
$F_{i}$ which are assumed to satisfy certain compatibility conditions. The
essential one is that the operators from a partition of unity, or rather a
partition of the identity operator I in the chosen Hilbert space. Here
(Definition \ref{Def1.0}) such a systems ($F_{i}$) are known as column
isometries. An extreme case of this is when a certain Cuntz relation
(Definition \ref{Def1.0}) is satisfied by. Referring back to our IFS
application, the extreme case of the operator relations turn out to correspond
to the limiting case of non-overlap. Using this operator theory, in this paper
we explore the fractal measures associated with the inherent self similarity
affine fractals, a subject involving both iterated function systems (IFSs),
and an aspect of quantum communication.

In the paper we aim to draw up connections between the following three areas,
quantum channels, fractal measures, and wavelets. Our understanding of quantum
measurement follows the tradition of Kraus \cite{Kra74}: By a quantum
measurement of a system refers to a system in some state and we want to
determine whether it has some property $E$, where $E$ should be thought of as
an element in a logic lattice of quantum yes-no questions. Measurement means
submitting the system to some procedure to determine whether the state
satisfies the property. The reference to system state in turn must be given an
operational meaning by reference to a statistical ensemble of systems. Each
measurement yields some definite value 0 or 1.

The connections to fractal measures and wavelets (see e.g., \cite{JoKr03})
derive from a common mathematical core based in turn on hierarchical patterns
common to them. Wavelets are computational bases in Hilbert space which make
use of the scaling law, with each step in the scaling creating a refinement in
a resolution of data, and taking advantage of similarity from one level of
resolution to the next. Hence we arrive at the characteristic feature of
fractals, see \cite{BEH89}, \cite{BHS05}.

Mathematically, an identification of quantum channels may be made with a
finite system of operators ($F_{i}$) which together form a column isomety; see
(1) below. In the physics literature, they are known as Kraus-systems, and
mathematically they generalize the so called Cuntz relations; see
\cite{BEH89}, \cite{BHS05}, \cite{DuJo05}, \cite{ShWe49}, \cite{Cun77},
\cite{FNW92}, \cite{Kra74}. The systems ($F_{i}$) which are the focus of our
paper in turn determine completely positive mappings $\alpha$ as recalled
below. In this paper we show that repeated application of operator system
yields the kind of refinement that underlies both wavelets and fractals, and
moreover that the statistical properties from quantum channels is of
significance in decomposition problems from wavelets. And with the scaling
from wavelet theory now corresponding to $\alpha$.

Both the quantum mechanical measurement problem and IFSs have as starting
point a finite set of operations: in the case of IFSs they are geometric, and
in the quantum case, they involve channels of Hilbert spaces and associated
operator systems. The particular aspects of IFSs we have in mind are studied
in \cite{DuJo06b}; and the relevant results from quantum communication in
\cite{KLPL06} and \cite{Kri05}. We begin the Introduction with some background
and motivation on IFSs. The operator theory, the fractal measures and their
applications are then taken up more systematically in section II below.

Let $\mathcal{H}$ be complex Hilbert space, and let $A$ be a finite set. We
will be interested in an indexed set of operators $\left\{  F_{i}|i\in
A\right\}  $ satisfying%
\begin{equation}
\sum_{i\in A}F_{i}^{\ast}F_{i}=I \label{Eq1}%
\end{equation}
where $I$ denotes the identity operator in $\mathcal{H}$.

\begin{definition}
\label{Def1.0}A finite system of operators $F_{i}$ in a Hilbert space
$\mathcal{H}$ is said to be a column isometry if
\[
\mathcal{H}\ni\xi\rightarrow\left(
\begin{array}
[c]{c}%
F_{1}\xi\\
\vdots\\
F_{N}\xi
\end{array}
\right)  \in\mathcal{H}^{\oplus N}%
\]
is isometric; and it is said to be a Cuntz-system if also
\[
F_{i}F_{j}^{\ast}=S_{i,j}I\text{.}%
\]

\end{definition}

Because of a certain reasoning outlined in the references below such systems
are called \textit{measurements} in quantum probability; see e.g.,
\cite{JoKr03}, \cite[OA0404553]{Kri05}, or \cite{Wer00}, \cite[quant-ph
101061]{Wer01}, \cite{ScWe01}.But they arise in other fields as well, in
representation theory, in geometric measure theory, and in wavelet analysis;
see e.g., \cite{DuJo06}, \cite{Jor06}, \cite{Jor05}, \cite{Jor03}, and
\cite[Minimality, Adv. Math]{Jor01}. Closely related systems of Hilbert space
operators play a big role in the theory of frames \cite{FJLO04},
\cite{CFKLT06} and their engineering applications. Suppose $\#A=N$. Denote by
$\mathbf{N}$ the cyclic group of order $N$ viz., $\mathbb{Z}/N\mathbb{Z}%
\cong\left\{  0,1,\cdots,N-1\right\}  $ or $\mathbb{Z}_{N},$ and set
\begin{equation}
\Omega\text{:}=\mathbf{N}^{\mathbb{N}\,}=\left\{  \text{all functions
}\mathbb{N}\rightarrow\mathbb{Z}_{N}\right\}  \text{.} \label{Eq2}%
\end{equation}
We shall give $\Omega$ its Tychonoff topology, and we view it as a compact
Hausdorff space.

Let $\mathcal{F}=\left(  F_{i}\right)  _{i\in\mathbf{N}}$ be a measurement,
and let $\psi\in\mathcal{H}$ be a unit-vector, i.e., a quantum mechanical
\textit{pure state}. Then it is immediate that
\begin{equation}
\mu_{\psi}\left(  i\right)  \text{:}=\left\Vert F_{i}\psi\right\Vert
^{2},~i\in\mathbf{N} \label{Eq3}%
\end{equation}
is a probability distribution on $\mathbf{N}$.

The measure in the title of the present paper refers operator-valued, or
scalar-valued measures on $\Omega=\Omega_{N}$ \textit{induced} by (\ref{Eq1})
-- (\ref{Eq3}). the induction from $\mathbb{Z}_{N}$ to $\Omega_{N}$ is based
on the Kolmogorov consistency condition \cite{Kol77}$,$ \cite{Jor06}, as follows:

\begin{definition}
\label{Def1.1}\ \medskip\newline\emph{(a)} Cylinder sets: Let $\left(
i_{1},\cdots,i_{K}\right)  \in\mathbb{Z}_{N}^{k}$, and set
\begin{equation}
C\left(  i_{1},\cdots,i_{k}\right)  =\left\{  w\in\Omega|w\left(  1\right)
=i_{1},\cdots,w\left(  k\right)  =i_{k}\right\}  \text{.} \label{Eq4}%
\end{equation}

\noindent\emph{(b)} Operator valued conditional probabilities:%
\begin{equation}
P\left(  C\left(  i_{1},\cdots,i_{k}\right)  \right)  =F_{i_{1}}^{\ast}\cdots
F_{i_{k}}^{\ast}F_{i_{k}}\cdots F_{i_{1}} \label{Eq5}%
\end{equation}

\noindent\emph{(c)} Kolmogorov consistency: Formula below.

Since we have the disjoint union
\begin{equation}
C\left(  i_{1},\cdots,i_{k}\right)  =\bigcup_{j\in\mathbb{Z}_{N}}C\left(
i_{1},\cdots,i_{k},j\right)  \text{,} \label{Eq7}%
\end{equation}
we need the formula
\begin{equation}
P\left(  C\left(  i_{1},\cdots,i_{k}\right)  \right)  =\sum_{j\in
\mathbb{Z}_{N}}P\left(  C\left(  i_{1},\cdots,i_{k},j\right)  \right)
\label{Eq1.7}%
\end{equation}
in order to extend \emph{(}\ref{Eq5}\emph{)} to a probability measure $P$
defined on the Borel subsets of $\Omega$. On the other hand, it is easy to see
that \emph{(}\ref{Eq7}\emph{)} is satisfied by \emph{(}\ref{Eq5}\emph{)}. Just
use the basic formula \emph{(}\ref{Eq1}\emph{)} for the given measurement
$\mathcal{F}$.
\end{definition}

\begin{lemma}
\label{Lem1.1}There is a unique positive operator-valued probability measure
$P$ defined on $\Omega$, and satisfying \emph{(}\ref{Eq5}\emph{)} For any
Borel set $B\subset\Omega$, $P\left(  B\right)  $ is well defined, and
$\langle\psi|P\left(  B\right)  \psi\rangle\geq0$ for all $\psi\in\mathcal{H}%
$. Moreover, $P$ is sigma-additive, or countably additive, i.e.,
\begin{equation}
P\left(  \bigcup\nolimits_{j=1}^{\infty}B_{j}\right)  =\sum_{j=1}^{\infty
}P\left(  B_{j}\right)  \label{Eq8}%
\end{equation}
whenever $B_{1},B_{2},\cdots$ are disjoint Borel sets;
\begin{equation}
P\left(  \Omega\right)  =I \label{Eq9}%
\end{equation}

\end{lemma}

\begin{proof}
The argument for the existence and uniqueness of the extension is a standard
application of Kolmogorov consistency. See \cite{Kol77}, \cite{Jor06},
\cite{Jor03}, \cite{Jor05} for more details.
\end{proof}

\begin{Examples}
\label{Exs1.1}~\medskip

\emph{(}a\emph{)} $N=2,~\mathcal{H}=\mathcal{C}^{2},~\psi=\left(
\genfrac{}{}{0pt}{}{1}{0}%
\right)  $, and
\begin{equation}
F_{0}=\frac{1}{\sqrt{2}}\left(
\begin{array}
[c]{ll}%
1 & 0\\
0 & 1
\end{array}
\right)  ,~F_{1}=\frac{1}{\sqrt{2}}\left(
\begin{array}
[c]{rr}%
1 & 0\\
0 & -1
\end{array}
\right)  \text{.} \label{Eq10}%
\end{equation}
Then
\begin{equation}
\mu_{\psi}\left(  C\left(  i_{1},\cdots,i_{k}\right)  \right)  =2^{-k}%
,~i_{j}\in\left\{  0,1\right\}  \text{.} \label{Eq11}%
\end{equation}
\emph{~}\medskip

\emph{(}b\emph{)} $N=3,~\mathcal{H}=\mathcal{C}^{3},~\psi=\left(
\begin{array}
[c]{l}%
1\\
0\\
0
\end{array}
\right)  $, and
\begin{equation}
F_{0}=\frac{1}{\sqrt{2}}\left(
\begin{array}
[c]{lll}%
1 & 0 & 0\\
0 & 1 & 0\\
0 & 0 & 0
\end{array}
\right)  ,~F_{1}=\left(
\begin{array}
[c]{lll}%
0 & 0 & 0\\
0 & 0 & 1\\
0 & 0 & 0
\end{array}
\right)  ,\text{~}F_{2}=\frac{1}{\sqrt{2}}\left(
\begin{array}
[c]{ccc}%
1 & 0 & 0\\
0 & -1 & 0\\
0 & 0 & 0
\end{array}
\right)  \text{.} \label{Eq12}%
\end{equation}
Then, for $i_{j}\in\left\{  0,1,2\right\}  ,~k\in\mathbb{N}$, and $1\leq j\leq
k$, we have
\begin{equation}
\mu_{\psi}\left(  C\left(  i_{1},\cdots,i_{k}\right)  \right)  =\left\{
\begin{array}
[c]{l}%
0\text{ if some }i_{j}=1\\
2^{-k}\text{ otherwise.}%
\end{array}
\right.  \label{Eq13}%
\end{equation}

\end{Examples}

\begin{proposition}
\label{Prop1.1}If we introduce $N$-adic partitions of the unit-interval
$\left[  0,1\right]  $ as follows
\begin{equation}
C\left(  i_{1},\cdots,i_{k}\right)  \rightarrow\left[  \frac{i_{1}}{N}%
+\cdots+\frac{i_{k}}{N^{k}},~\frac{i_{1}}{N}+\cdots+\frac{i_{k}}{N^{k}}%
+\frac{1}{N^{k}}\right)  \label{Eq14}%
\end{equation}
then the measures $P\left(  \cdot\right)  $ and $\mu_{\psi}\left(
\cdot\right)  =\langle\psi|P\left(  \cdot\right)  \psi\rangle$ are Borel
measures, each supported on $\left[  0,1\right]  $. If $N=2$, the measure
\emph{(}\ref{Eq11}\emph{)} in Example \ref{Exs1.1} \emph{(}a\emph{)} turns out
to be merely Lebesgue measure restricted to $\left[  0,1\right]  $. If $N=3$,
the measure \emph{(}\ref{Eq13}\emph{)} in Example \ref{Exs1.1} \emph{(}%
b\emph{)} is the middle-third Cantor measure supported in the Cantor set
$\mathbf{X}_{3}$.
\end{proposition}

\begin{remark}
\label{Rem1.1}Recall $\mathbf{X}_{3}$ is the unique \emph{(}compact\emph{)}
subset of $\mathbb{R}$ satisfying
\begin{equation}
3\mathbf{X}_{3}=\mathbf{X}_{3}\cup\left(  \mathbf{X}_{3}+2\right)  \text{,}
\label{Eq15}%
\end{equation}
and the Cantor measure $\mu=\mu_{3}$ is the unique Borel measures satisfying
\begin{equation}
\int f\left(  x\right)  ~d\mu\left(  x\right)  =\frac{1}{2}\left(  \int
f\left(  \frac{x}{3}\right)  ~d\mu\left(  x\right)  +\int f\left(  \frac
{x+2}{3}\right)  ~d\mu\left(  x\right)  \right)  \label{Eq16}%
\end{equation}
for all bounded Borel functions $f$.
\end{remark}

\begin{proof}
\textbf{(of Proposition I.4)} The assertions follow from standard applications
of the Kolmogorov extension principle, and the reader is referred to
\cite{Jor03}, \cite{Jor05} for additional discussion.
\end{proof}

The column isometries ($F_{i}$) introduced above can be viewed as Kraus
operators \cite{Kra74} from the theory of quantum channels, modeling
\textquotedblleft instruments\textquotedblright\ in Kraus's formulation, and
operating on quantum systems by producing a classical measurement (outcome.)
The theory of finitely correlated states \cite{FNW92} has an instrument
generating a classical state which takes the form of a measure on an infinite
product of a finite alphabet; a construction which parallels the theme of our
paper. It would be intriguing to explore how the physics of \cite{FNW92}
reflects itself in the measures more directly associated to fractal measures,
and wavelets. Part of the answer lies in how families of wavelet packets adapt
to signals (or in 2D) to images, see e.g., \cite{JoKr03}. The operators in the
system ($F_{i}$) are iterated in steps with each iteration step creating a
subdivision of the masks in the previous more coarse resolution. A monomial of
degree k in the generators $F_{i}$ corresponds to a $k$ fold subdivision. Even
though this algorithm is \textquotedblleft classical,\textquotedblright\ the
generators $F_{i}$ are non-commuting operators. The limit of this iterative
scheme as $k$ tends to infinity is made precise by the measures that we
analyze in theorems II.3 and III.2 below.

\section{Fractal Scales}

The authors of \cite{DuJo06a} recently adapted the discrete wavelet algorithms
to fractals, and the present work extends \cite{DuJo06a}.

The distinction between the two prototypical cases (a) and (b) in Example
\ref{Exs1.1} can be made precise in a number of different ways; for example,
it can be checked that the fractal dimension (in this case = the Hausdorff
dimension) of (a) is $1$, and of (b) it is $s=\frac{\ln2}{\ln3}=\log_{3}(2)$.
For our present discussion, the following definition of the fractal dimension
will suffice: If a subset $\mathbf{X}\subset\mathbb{R}$\thinspace$^{d}$ is
obtained by the iteration of a finite family of contractive and affine maps
$\mathbb{R}$\thinspace$^{d}\rightarrow\mathbb{R}$\thinspace$^{d}$ then the
fractal dimension $s$ of $\mathbf{X}$
\begin{equation}
s=\frac{\log\left(  \text{\textit{number of replicas}}\right)  }{\log\left(
\text{\textit{magnification factor}}\right)  }\text{.} \label{Eq17}%
\end{equation}

Following Proposition \ref{Prop1.1}, especially (\ref{Eq14}), we will restrict
attention in the following to subsets of $\left[  0,1\right]  $ and measures
defined on the Borel subsets of $\left[  0,1\right]  $. If $J\subset\left[
0,1\right]  $ is a subinterval, we denote by $\left\vert J\right\vert $ the
length of $J$. \bigskip

\begin{definition}
\label{Def2.1}Let $\mu$ be a probability measure on $\left[  0,1\right]  $
defined on the Borel sets.
\end{definition}

We say that $s_{-}$ is a \textit{lower scale} of $\mu$ if
\begin{equation}
\underset{\left\vert J\right\vert \rightarrow0}{\underset{\mu\left(  J\right)
>0}{\lim\inf}}\frac{\mu\left(  J\right)  }{\left\vert J\right\vert ^{s_{-}}%
}>0\text{;} \label{Eq18}%
\end{equation}
and we say that $s_{+}$ is an \textit{upper scale} of $\mu$ if
\begin{equation}
\underset{\left\vert J\right\vert \rightarrow0}{\underset{\mu\left(  J\right)
>0}{\lim\sup}}\frac{\mu\left(  J\right)  }{\left\vert J\right\vert ^{s_{+}}%
}<\infty\text{.} \label{Eq19}%
\end{equation}
It is easy to see that the Cantor measure $\mu=\mu_{\psi}$ in Example
\ref{Exs1.1} (b) has both upper and lower scale $s_{+}=s_{-}=s=\log_{3}\left(
2\right)  $.

Our next result is motivated by examples from wavelet analysis. Before stating
our general result we first recall the wavelet examples. To emphasize our
point, we do not consider the wavelet examples in the widest generality.

\begin{example}
\label{Ex2.1}{\bfseries{\upshape Discrete Wavelet Transforms.}} Let
$a_{0},a_{1},\cdots$ be a sequence of complex numbers such that
\begin{equation}
\sum_{j}\overline{a_{j}}a_{j+2k}=\delta_{0,k}\text{.} \label{Eq20}%
\end{equation}
In the summation \emph{(}\ref{Eq20}\emph{)}, it is understood that terms are
zero if the subindex is not in the range where $a\not =0$.
\end{example}

We define operators $F_{0}$ and $F_{1}$ on the Hilbert space $\mathcal{H}%
$=$\ell^{2}$ as follows:
\begin{equation}
\left(  F_{0}\xi\right)  _{j}\text{:}=\sum_{k}a_{2j-k}\xi_{k} \label{Eq21}%
\end{equation}
and
\begin{equation}
\left(  F_{1}\xi\right)  _{j}\text{:}=\sum_{k}\left(  -1\right)  ^{k}%
\overline{a_{1-2j+k}}\xi_{k}\text{.} \label{Eq22}%
\end{equation}
\bigskip Then it is easy to check that (\ref{Eq1}) holds, and so the pair
$(F_{0},F_{1})$ defines a measurement in the sense of the definition in
Section I. In this case, more is true: The adjoint operators $F_{i}^{\ast}$
are isometries with orthogonal ranges, i.e.,
\begin{equation}
F_{i}F_{j}^{\ast}=S_{i,j}I\text{.} \label{Eq23}%
\end{equation}
If the sequence $a_{0},a_{1},\cdots$ from (\ref{Eq20}) is finite, then it is
easy to see that the number of non-zero terms must necessarily be even. We
consider $2D$ scalars,
\begin{equation}
a_{0},a_{1},\cdots,a_{2D-1}\text{,} \label{Eq24}%
\end{equation}
and the corresponding two $\left(  2D-1\right)  $ by $\left(  2D-1\right)  $
matrices $F_{0}$ and $F_{1}$ defined as follows:
\begin{equation}
F_{0}=\left(
\begin{array}
[c]{cccccccc}%
a_{0} & 0 & 0 & \cdots & \cdots & \cdots & 0 & 0\\
a_{2} & a_{1} & a_{0} &  &  &  & \vdots & \vdots\\
\vdots & a_{3} & a_{2} &  &  &  & \vdots & \vdots\\
\vdots & \vdots &  &  &  &  & 0 & 0\\
a_{2D-2} & \vdots &  &  &  & a_{1} & 0 & 0\\
0 & a_{2D-1} & a_{2D-2} &  &  & a_{2} & a_{1} & a_{0}\\
0 & 0 & 0 &  &  & \vdots & a_{3} & a_{2}\\
\vdots & \vdots & \vdots &  &  & a_{2D-2} & \vdots & \vdots\\
0 & 0 & 0 & \cdots & 0 & 0 & a_{2D-1} & a_{2D-2}%
\end{array}
\right)  \label{Eq25}%
\end{equation}
and $F_{1}$ built the same way, but using the numbers
\begin{equation}
b_{k}\text{:}=\left(  -1\right)  ^{k}\overline{a_{2D-1-k}} \label{Eq26}%
\end{equation}
For $D=2$, the two matrices are simply
\begin{equation}
F_{0}=\left(
\begin{array}
[c]{rrr}%
a_{0} & 0 & 0\\
a_{2} & a_{1} & a_{1}\\
0 & a_{3} & a_{2}%
\end{array}
\right)  \text{ and }F_{1}=\left(
\begin{array}
[c]{rrr}%
\overline{a_{3}} & 0 & 0\\
\overline{a_{1}} & -\overline{a_{2}} & \overline{a_{3}}\\
0 & -\overline{a_{0}} & \overline{a_{1}}%
\end{array}
\right)  \label{Eq27}%
\end{equation}
Staying with $a_{0},a_{1},a_{2},a_{3}$, there are practical reasons in wavelet
analysis to add the following two requirements to (\ref{Eq20}):\medskip

(i) $a_{i}\in\mathbb{R}$\thinspace\newline and

(ii) $\sum\limits_{i=0}^{3}a_{i}=\sqrt{2}$.\medskip

\noindent Taking the combined conditions together, it can be shown that
$a_{0},a_{1},a_{2},a_{3}$ are determined by a single real parameter $\beta$
thus;
\begin{equation}
\left\{
\begin{array}
[c]{l}%
a_{0}=\frac{1}{2\sqrt{2}}\left(  1+\sqrt{2}\cos\beta\right)  \\
a_{1}=\frac{1}{2\sqrt{2}}\left(  1+\sqrt{2}\sin\beta\right)  \\
a_{2}=\frac{1}{2\sqrt{2}}\left(  1-\sqrt{2}\cos\beta\right)  \\
a_{3}=\frac{1}{2\sqrt{2}}\left(  1-\sqrt{2}\sin\beta\right)  \text{.}%
\end{array}
\right.  \label{Eq28}%
\end{equation}
A consequence of (\ref{Eq28}) is that each of the three pairs $\left(
a_{0},a_{1}\right)  $, $\left(  a_{0},a_{3}\right)  $, and $\left(
a_{1},a_{2}\right)  $ lies on the circle
\begin{equation}
\left(  x-\frac{1}{2\sqrt{2}}\right)  ^{2}+\left(  y-\frac{1}{2\sqrt{2}%
}\right)  ^{2}=\frac{1}{4}\text{;}\label{Eq29}%
\end{equation}
see Fig. 1.%
\[%
{\parbox[b]{3.7775in}{\begin{center}
\includegraphics[
natheight=3.273500in,
natwidth=4.319600in,
height=3.0295in,
width=3.7775in
]%
{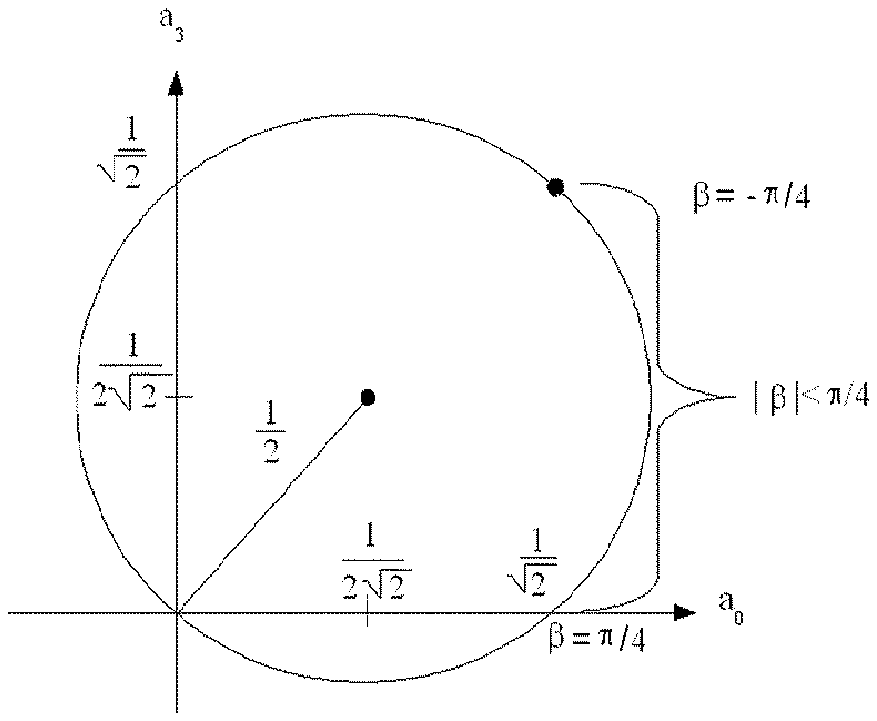}%
\\
Figure 1. One of the three pairs $(a_0,a_3)$
\end{center}}}%
\]

\begin{theorem}
\label{Theo2.1}Let the numbers $a_{0},a_{n},\cdots,a_{2D-1}$ be given, and
suppose \emph{(}\ref{Eq20}\emph{)} is satisfied. Let $F_{0}$ and $F_{1}$ be
the corresponding matrices determined by \emph{(}\ref{Eq25}\emph{)}%
--\emph{(}\ref{Eq26}\emph{)}. Suppose further that $a_{0}\cdot a_{2D-1}%
\not =0$. Let
\begin{equation}
\alpha\text{\emph{:}}=\max\left(  \left\vert a_{0}\right\vert ^{2},\left\vert
a_{2D-1}\right\vert ^{2}\right)  \text{.} \label{Eq30}%
\end{equation}
Then the number
\begin{equation}
s=\log_{2}\alpha^{-1}=-\frac{\ln\alpha}{\ln2} \label{Eq31}%
\end{equation}
is a lower scale of $\mu_{0}\left(  \cdot\right)  =\langle e_{0}|P\left(
\cdot\right)  e_{0}\rangle$ where
\begin{equation}
e_{0}=\left(
\begin{array}
[c]{l}%
1\\
0\\
\vdots\\
0
\end{array}
\right)  \text{.} \label{Eq32}%
\end{equation}

It follows in particular that if $\alpha>\frac{1}{2}$, then $\mu_{0}$ has a
lower scale $s>1$.

Moreover, we can get the lower estimate \emph{(}\ref{Eq18}\emph{)} satisfied
for dyadic intervals $J$ inside any non-empty open subset of $\left[
0,1\right]  $.
\end{theorem}

\begin{proof}
From the representation (\ref{Eq25}) of the two matrices $F_{0}$ and $F_{1}$
we conclude that
\begin{equation}
F_{0}^{\ast}e_{0}=\overline{a_{0}}e_{0}\text{ and }F_{1}^{\ast}e_{0}%
=a_{2D-1}e_{0}\text{.} \label{Eq33}%
\end{equation}
With the dyadic representation (\ref{Eq14}), $N=2$, set
\begin{equation}
\xi=\frac{i_{1}}{2}+\cdots+\frac{i_{k}}{2^{k}} \label{Eq34}%
\end{equation}
and
\begin{equation}
F_{\xi}=F_{i_{k}}\cdots F_{i_{1}}\text{.} \label{Eq35}%
\end{equation}
Then it follows from (\ref{Eq15}) and Lemma \ref{Lem1.1} that
\begin{align*}
\mu_{0}\left(  \left[  \xi,\xi+2^{-k}\right)  \right)   &  =\langle
e_{0}|F_{i_{1}}^{\ast}\cdots F_{i_{k}}^{\ast}F_{i_{k}}\cdots F_{i_{1}}%
e_{0}\rangle\\
&  =\langle F_{i_{k}}\cdots F_{i_{1}}e_{0}|F_{i_{k}}\cdots F_{i_{1}}%
e_{0}\rangle\\
&  =\left\Vert F_{\xi}e_{0}\right\Vert ^{2}\\
&  \geq\left\vert \langle e_{0}|F_{\xi}e_{0}\rangle\right\vert ^{2}\\
&  \text{{\tiny (Schwarz)}}\\
&  =\left\vert \langle F_{\xi}^{\ast}e_{0}|e_{0}\rangle\right\vert ^{2}\\
&  =\left\vert a_{0}\right\vert ^{2\cdot\#\left(  i=0\right)  }\cdot\left\vert
a_{2D-1}\right\vert ^{2\cdot\#\left(  i=1\right)  }\\
&  \text{{\tiny using (\ref{Eq34})}}%
\end{align*}

Let $V$ be a non-empty open subset of $\left[  0,1\right]  $, and pick
$k\in\mathbb{N}$\thinspace$and$ $\xi$ as in (\ref{Eq34}) such that the
interval $\left[  \xi,\xi+2^{-k}\right)  $ is contained in $V$.

We now turn to the two possibilities for the number $\alpha$ in (\ref{Eq30}).

If $\alpha=\left\vert a_{0}\right\vert ^{2}$, then
\begin{align*}
\mu_{0}\left(  \left[  \xi,\xi+2^{-k-n}\right)  \right)   &  =\left\Vert
F_{0}^{n}F_{\xi}e_{0}\right\Vert ^{2}\\
&  \geq\alpha^{\#\left(  i=0\right)  +n}\left\vert a_{2D-1}\right\vert
^{2\cdot\#\left(  i=1\right)  }%
\end{align*}
and we conclude that the expression
\begin{equation}
\frac{\mu_{0}\left(  \left[  \xi,\xi+2^{-k-n}\right)  \right)  }{2^{-s\left(
k+n\right)  }} \label{Eq36}%
\end{equation}
is bounded below as $n\rightarrow\infty$, and hence (\ref{Eq18}) holds for
$s=\log_{2}\left(  \alpha^{-1}\right)  $, see (\ref{Eq31}).

If instead $\alpha=\left\vert a_{2D-1}\right\vert ^{2}$, then
\[
\mu_{0}\left(  \left[  \xi+2^{-k}\left(  1-2\right)  ^{-n}\right)  ,\xi
+2^{-k}\right)  \geq\left\vert a_{0}\right\vert ^{2\#\left(  i=0\right)
}\alpha^{\#\left(  i=1\right)  +n}%
\]
by the same reasoning used in the first case. We now get the lower estimate
(\ref{Eq18}) satisfied for the intervals $J=\left[  \xi+2^{-k}\left(
1-2^{-n}\right)  ,\xi+2^{-k}\right)  $ as $n\rightarrow\infty$. This completes
the proof
\end{proof}

\section{Upper and Lower Fractal Scales for the Measure $\mu_{0}$}

Consider the example outlined in (\ref{Eq28}) above. The two matrices $F_{0}$
and $F_{1}$ are used in wavelet analysis where they refer to low-pass and
high-pass filters; terms that derive from signal processing, see \cite{Dau92}
and \cite{BrJo02b}.

Recall that when $a_{0},a_{1},a_{2},a_{3}$ are given by (\ref{Eq28}) then
there are solutions $\phi,\psi$ in $L^{2}\left(  \mathbb{R}\,\right)  $ to
\[
\left\{
\begin{array}
[c]{l}%
\phi\left(  x\right)  =\sqrt{2}\left(  a_{0}\phi\left(  2x\right)  +a_{1}%
\phi\left(  2x-1\right)  +a_{2}\phi\left(  2x-2\right)  +a_{3}\phi\left(
2x-3\right)  \right) \\
\quad\int_{\mathbb{R}\,}\phi\left(  x\right)  ~dx=1\\
\psi\left(  x\right)  =\sqrt{2}\left(  a_{3}\phi\left(  2x\right)  -a_{2}%
\phi\left(  2x-1\right)  +a_{1}\phi\left(  2x-2\right)  -a_{0}\phi\left(
2x-3\right)  \right) \\
\quad\int_{\mathbb{R}\,}\psi\left(  x\right)  ~dx=0\text{,}%
\end{array}
\right.
\]
and when $\beta\in\mathbb{R}$\thinspace$\diagdown\left(  \left\{  \pm\frac
{\pi}{4},\pm\frac{3\pi}{4}\right\}  +\mathbb{Z}2\pi\right)  $, then the two
functions $\phi$ (the scaling function) and $\psi$ (the wavelet) satisfy the
further conditions

$%
\begin{array}
[c]{l}%
\int\limits_{\mathbb{R}\,}\phi\left(  x\right)  \phi\left(  x-k\right)
~dx=\delta_{0,k},~k\in\mathbb{Z}\,
\end{array}
$\newline and

$%
\begin{array}
[c]{l}%
\left\{  2^{j/2}\psi\left(  2^{j}x-k\right)  |j,k\in\mathbb{Z}\,\right\}
\end{array}
$\newline is an orthonormal basis for the Hilbert space $L^{2}\left(
\mathbb{R}\,\right)  $.

The general feature of the matrices (\ref{Eq27}) is the slanted shape; and the
$F_{i}$s are called \textit{slanted Toeplitz matrices}.

\begin{example}
\label{Ex3.1}{{\bfseries{\upshape A one-parameter family of wavelets.}} For
the case }$D=2$ treated in \emph{(}\ref{Eq28}\emph{)}, we have
\begin{equation}
\left(  a_{0}-\frac{1}{2\sqrt{2}}\right)  ^{2}+\left(  a_{3}-\frac{1}%
{2\sqrt{2}}\right)  ^{2}=\frac{1}{4} \label{Eq37}%
\end{equation}
and it follows that the number%
\begin{equation}
\alpha=\max\left(  a_{0}^{2},a_{3}^{2}\right)  \label{Eq38}%
\end{equation}
satisfies $\alpha>\frac{1}{2}$ when the parameter $\beta$ is in one of the two
intervals\emph{:}

\emph{(}i\emph{)} $\left\vert \beta\right\vert \,\frac{\pi}{4}$ where
$\alpha=a_{0}^{2}$;\newline or

\emph{(}ii\emph{)} $-\frac{3\pi}{4}<\beta<-\frac{\pi}{4}$ where $\alpha
=a_{3}^{2}$;

\noindent see Fig. 2 below. In these two regions the scale number $s$
satisfies $s>1$; hence the fractal feature of the measure $\mu_{0}$.
\end{example}

\[%
{\parbox[b]{4.7729in}{\begin{center}
\includegraphics[
trim=0.000000in 0.000000in 0.066886in 0.000000in,
height=2.494in,
width=4.7729in
]%
{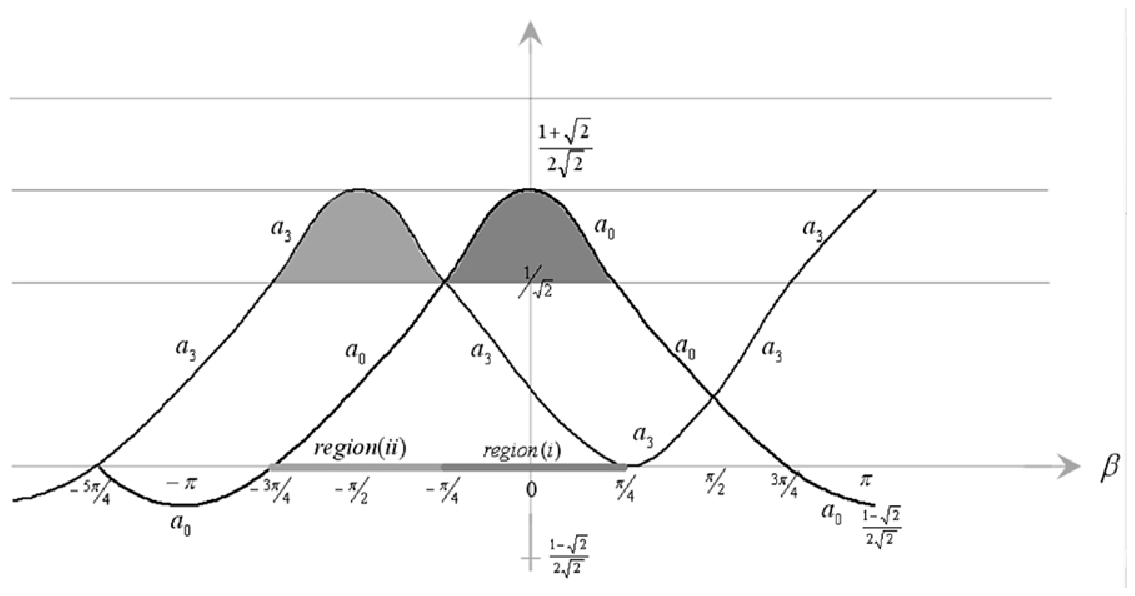}%
\\
Figure 2. The two functions $a_0=a_0\left(  \beta\right)  $ and $%
a_3=a_3\left(  \beta\right)  $, (\ref{Eq28}). Two regions (i) and (ii) with
scale number $s$ satisfying $s>1$.
\end{center}}}%
\]

When the four numbers $a_{0},a_{1},a_{2},a_{3}$ that make up the matrix
$F_{0}$ are given by formulas (\ref{Eq28}), one easily computes the spectrum
of $F_{0}$ as follows:%
\begin{equation}
\operatorname*{spec}\left(  F_{0}\right)  =\left\{  a_{0}\left(  \beta\right)
,\frac{1}{\sqrt{2}},\frac{\sin\beta-\cos\beta}{2}\right\}  \text{,}%
\label{Eq39}%
\end{equation}
and we have sketched the point
\begin{equation}
\lambda\left(  \beta\right)  =\frac{\sin\beta-\cos\beta}{2}\label{Eq40}%
\end{equation}
in the spectrum in Figure 3. An inspection shows that $\lambda\left(
\beta\right)  $ is not dominant in $\operatorname*{spec}\left(  F_{0}\right)
$ in the sense that the inequalities
\begin{equation}
a_{0}\left(  \beta\right)  >\frac{1}{\sqrt{2}}>\left\vert \lambda\left(
\beta\right)  \right\vert \label{Eq41}%
\end{equation}
hold in the region (i) from Figure 2.%
\[%
{\parbox[b]{5.408in}{\begin{center}
\includegraphics[
height=2.2349in,
width=5.408in
]%
{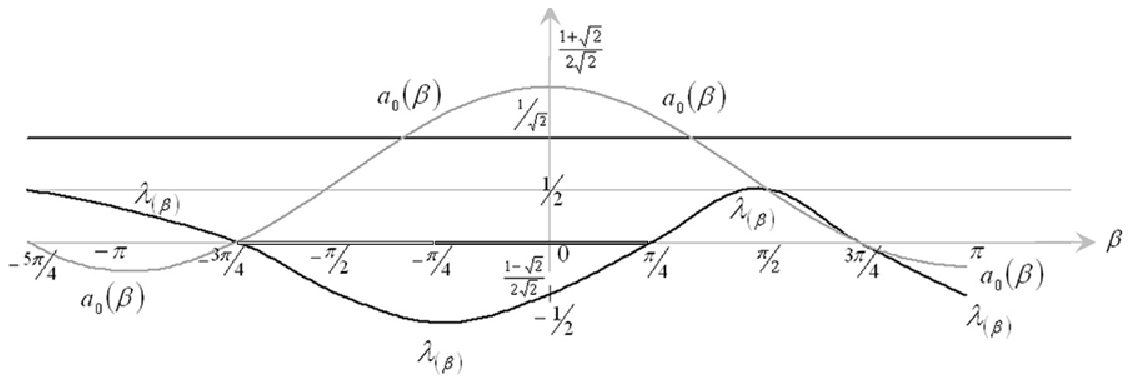}%
\\
The eigenvalue $\lambda\left(  \beta\right)  =\frac{\sin\beta-\cos\beta}{2}$
is not dominant. (i):$a_0\left(  \beta\right)  >\frac{1}{\sqrt{2}}%
>\lambda\left(  \beta\right)  ^2$
\end{center}}}%
\]

We will now turn to our analysis of the two-sided scale bound for the measure
$\mu_{0}$ and we show how it applies to the matrices $F_{0}$ and $F_{1}$ which
are used in wavelet theory.

\begin{theorem}
\label{Theo3.1}Let the numbers $a_{0},a_{1},\cdots,a_{2D-1}$ satisfy condition
\emph{(}\ref{Eq20}\emph{)}, and in addition
\begin{equation}
\sum_{j=0}^{2D-1}a_{j}=\sqrt{2\text{.}} \label{Eq42}%
\end{equation}
The two matrices $F_{0}$ and $F_{1}$ are defined as in \emph{(}\ref{Eq25}%
\emph{)}--\emph{(}\ref{Eq26}\emph{)}. We make the following additional
assumptions on the spectrum of $F_{0}$:
\begin{equation}%
\begin{array}
[c]{ll}%
\!\!\!\emph{(}i\emph{)}\, & a_{0}\cdot a_{2D-1}\not =0;\qquad\qquad
\qquad\text{\qquad\qquad\qquad\qquad\qquad\qquad\qquad}\qquad\qquad\qquad\;\;
\end{array}
\label{Eq43}%
\end{equation}

\begin{equation}%
\begin{array}
[c]{ll}%
\!\!\!\emph{(}ii\emph{)}\, & \left\vert a_{0}\right\vert >\max\left\{
\left\vert \lambda\right\vert \lambda\in\operatorname*{spec}\left(
F_{0}\right)  \diagdown\left\{  a_{0}\right\}  \right\}  ;\qquad
\text{\qquad\qquad\qquad\qquad\qquad}\qquad\quad\quad\;
\end{array}
\label{Eq44}%
\end{equation}

\begin{equation}%
\begin{array}
[c]{ll}%
\!\!\!\emph{(}iii\emph{)} & \text{the algebraic multiplicity of }a_{0}\text{
in}\operatorname*{spec}\left(  F_{0}\right)  \text{ is one.\qquad\qquad
\qquad\qquad\qquad\ }%
\end{array}
\label{Eq45}%
\end{equation}

Then there is a unique vector $v$ such that
\begin{equation}
F_{0}v=a_{0}v\text{ and }\langle e_{0}|v\rangle=1 \label{Eq46}%
\end{equation}
where $e_{0}$ is the vector \emph{(}\ref{Eq32}\emph{)} with $1$ is the first
slot and zeros in the rest.

Moreover
\begin{equation}
s=\log_{2}\left(  \left\vert a_{0}\right\vert ^{-1}\right)  =-\frac
{2\ln\left\vert a_{0}\right\vert }{\ln2} \label{Eq47}%
\end{equation}
is both an upper scale and a lower scale in every non-empty open subset of
$\left[  0,1\right]  $ for the measure
\begin{equation}
\mu_{0}\left(  \cdot\right)  =\langle e_{0}|P\left(  \cdot\right)
e_{0}\rangle\text{.} \label{Eq48}%
\end{equation}

\end{theorem}

\begin{proof}
Set $w=\underset{2D-1~\text{ones}}{(\underbrace{1,1,\cdots,1}})$. Then
(\ref{Eq42}) implies
\begin{equation}
wF_{0}=\frac{1}{\sqrt{2}}w\text{ or equivalently }F_{0}^{\ast}w^{\ast}%
=\frac{1}{\sqrt{2}}w^{\ast} \label{Eq49}%
\end{equation}
where $w^{\ast}$ denotes the column vector corresponding to $w$. So property
(\ref{Eq44}), i.e., (ii) above, yields%
\begin{equation}
\left\vert a_{0}\right\vert ^{2}>\frac{1}{2}\text{ and }\left\vert
a_{0}\right\vert >\left\vert a_{2D-1}\right\vert \text{.} \label{Eq50}%
\end{equation}
Since $F_{0}^{\ast}e_{0}=\bar{a}_{0}e_{0}$, (\ref{Eq45}) implies that $a_{0}$
is not in the spectrum of the matrix $G$ arising from $F_{0}$ by deletion of
the first row and the first column, i.e.,
\begin{equation}
G=\left(
\begin{array}
[c]{lllllllll}%
a_{1} & a_{0} & 0 & 0 & \cdots & \cdots & \cdots & 0 & 0\\
a_{3} & a_{2} & a_{1} & a_{0} &  &  &  & \vdots & \vdots\\
\vdots & \vdots & \vdots & \vdots &  &  &  & \vdots & \vdots\\
a_{2D-1} & a_{2D-2} & a_{2D-1} & a_{2D-1} &  &  &  & 0 & 0\\
0 & 0 &  &  &  &  &  & a_{1} & a_{0}\\
\vdots &  &  &  &  &  &  & a_{3} & a_{2}\\
\vdots &  &  &  &  &  &  & \vdots & \vdots\\
\vdots &  &  &  &  &  &  & \vdots & \vdots\\
0 & 0 & \cdots & \cdots & \cdots & \cdots & 0 & a_{2D-1} & a_{2D-2}%
\end{array}
\right)  \text{.} \label{Eq51}%
\end{equation}
Now define a vector $v$ in $\mathbb{C}^{2D-1}$ by
\begin{equation}
v=e_{0}+\left(  a_{0}I_{2D-2}-G\right)  ^{-1}\left(
\begin{array}
[c]{l}%
a_{2}\\
a_{4}\\
\vdots\\
a_{2D-2}\\
0\\
\vdots\\
0
\end{array}
\right)  \text{.} \label{Eq52}%
\end{equation}
To better visualize (\ref{Eq51}) the reader may check that, if $D=2$,
\begin{equation}
F_{0}\left(
\begin{array}
[c]{lll}%
a_{0} & 0 & 0\\
a_{2} & a_{1} & a_{0}\\
0 & a_{3} & a_{2}%
\end{array}
\right)  ,~G=\left(
\begin{array}
[c]{ll}%
a_{1} & a_{0}\\
a_{3} & a_{2}%
\end{array}
\right)  , \label{Eq53}%
\end{equation}
and
\begin{equation}
v=\left(  \left(  a_{0}I_{2}-\left(
\begin{array}
[c]{ll}%
a_{1} & a_{0}\\
a_{3} & a_{2}%
\end{array}
\right)  \right)  ^{-1}\left(
\begin{array}
[c]{l}%
a_{2}\\
0
\end{array}
\right)  \right)  =\left(
\begin{array}
[c]{c}%
1\\
\frac{\left(  a_{0}-a_{2}\right)  a_{2}}{p\left(  a_{0}\right)  }\\
\frac{a_{3}a_{2}}{p\left(  a_{0}\right)  }%
\end{array}
\right)  \label{Eq54}%
\end{equation}
where
\[
p\left(  a_{0}\right)  =a_{0}^{2}-\left(  a_{1}+a_{2}\right)  a_{0}+a_{1}%
a_{2}-a_{0}a_{3}\text{.}%
\]
Recall that the characteristic polynomial of $G$ is
\begin{equation}
p\left(  \lambda\right)  =\lambda^{2}-\left(  \operatorname*{trace}G\right)
\lambda+\det G\text{,} \label{Eq55}%
\end{equation}
and that $p\left(  a_{0}\right)  $ in the fraction of (\ref{Eq54}) is
evaluation of (\ref{Eq55}) at $\lambda=a_{0}$. Hence, assumption (\ref{Eq45})
comes into play.

Returning to the general case, we claim that $v$ satisfies (\ref{Eq46}).
Indeed, let $v$ be given by (\ref{Eq52}). then
\begin{align*}
Fv  &  =a_{0}e_{0}+\left(  I+G\left(  a_{0}I-G\right)  ^{-1}\right)  \left(
\begin{array}
[c]{l}%
a_{2}\\
a_{4}\\
\vdots\\
a_{2D-2}\\
0\\
\vdots\\
0
\end{array}
\right) \\
&  =a_{0}e_{0}+a_{0}\left(  a_{0}I-G\right)  ^{-1}\left(
\begin{array}
[c]{l}%
a_{2}\\
a_{4}\\
\vdots\\
a_{2D-2}\\
0\\
\vdots\\
0
\end{array}
\right) \\
&  =a_{0}v\text{,}\\
&  \text{and}\\
\langle e_{0}|v\rangle &  =1\text{.}%
\end{align*}
This proves (\ref{Eq46}). Conversely, using (\ref{Eq44})--(\ref{Eq45}), one
shows that the solution $v$ to (\ref{Eq46}) is unique.

Now let $V$ be a non-empty open subset of $\left[  0,1\right]  $. Then pick
$k\in\mathbb{N}$\thinspace and a dyadic fraction
\begin{equation}
\xi=\frac{i_{1}}{2}+\cdots+\frac{i_{k}}{2^{k}} \label{Eq56}%
\end{equation}
such that $\left[  \xi,\xi+2^{-k}\right)  \subset V$.

We now wish to estimate $\mu_{0}\left(  \left[  \xi,\xi+2^{-k-n}\right)
\right)  $ and get the asymptotic scaling rate as $n\rightarrow\infty$.

To that end, we prove in the next section (in a separate lemma; see especially
(\ref{Eq59}) that $\lim_{n\rightarrow\infty}a_{0}^{-n}F_{0}F_{\xi}%
e_{0}=\langle e_{0}|F_{\xi}e_{0}\rangle v=\langle F_{\xi}^{\ast}e_{0}%
|e_{0}\rangle v=a_{0}^{\#\left(  i=0\right)  }a_{2D-1}^{\#\left(  i=1\right)
}v$, and as a result $\lim\left\vert a_{0}\right\vert ^{-2n}\mu_{0}\left(
\left[  \xi,\xi+2^{-k-n}\right)  \right)  =\left\vert a_{0}\right\vert
^{2\cdot\#\left(  i=0\right)  }\left\vert a_{2D-1}\right\vert ^{2\cdot
\#\left(  i=1\right)  }\left\Vert v\right\Vert ^{2}n\rightarrow\infty$. Since
$\left\Vert v\right\Vert ^{2}\geq1$, the desired conclusion follows.
\end{proof}

In applications to wavelets, the measures in the title of the paper are used
in the computation of transition matrices for transformation between two
orthogonal families in the Hilbert space $L^{2}\left(  \mathbb{R}\right)
$:\smallskip

(i) a wavelet basis $\left(  2^{p/2}\psi\left(  2^{p}x-k\right)  \right)
_{p,k\in\mathbb{Z}}$;

\noindent and

(ii) a wavelet packet $\left(  \phi_{n}\right)  _{n\in\mathbb{N}_{0}%
,}\mathbb{N}_{0}=\left\{  0,1,2,\cdots\right\}  $;

$\quad\;\phi_{0}$:$=\phi$, $\phi_{1}$:$=\psi$,

$\quad\;\phi_{2n}\left(  x\right)  =\sqrt{2}\sum_{k\in\mathbb{Z}\,}a_{k}%
~\phi_{n}\left(  2x-k\right)  $,

$\quad\;\phi_{2n+1}\left(  x\right)  =\sqrt{2}\sum_{k\in\mathbb{Z}\,}%
b_{k}~\phi_{n}\left(  2x-k\right)  $.\smallskip

The adjustment of dyadic scaling in (i) is made with variations in
$p\in\mathbb{Z}$; and hence with the size of the dyadic intervals $J\left(
k,p\right)  =\left[  k2^{-p},~\left(  k+1\right)  2^{-p}\right)  $. The
concentration of mass at each $J\left(  k,p\right)  $ is determined by the measure.

\section{A Technical Lemma}

In the proof of Theorem \ref{Theo3.1} above, we relied on the following lemma
regarding operators in a finite-dimensional Hilbert space. While it is
analogous to the classical Perren-Frobenius theorem, our present result makes
no mention of positivity. In fact, our matrix entries will typically be complex.

\begin{notation}
\label{Not4.1}If $\mathcal{M}$ is a complex Hilbert space, we denote by
$L\left(  \mathcal{M}\right)  $ the algebra of all bounded linear operators on
$\mathcal{M}$. If $\mathcal{M}$ is also finite-dimensional, we will pick
suitable matrix representations for operators $F$\emph{:}$\mathcal{M}%
\rightarrow\mathcal{M}$.
\end{notation}

Suppose $\mathcal{M}$ contains two subspaces $\mathcal{M}_{i},$ $i=1,2$ such
that $\mathcal{M}_{1}\perp\mathcal{M}_{2}$ and $\mathcal{M}=\mathcal{M}%
_{1}\oplus\mathcal{M}_{2}$, then we get a block-matrix representation
\begin{equation}
F=\left(
\begin{array}
[c]{ll}%
A & B\\
C & D
\end{array}
\right)  \label{Eq57}%
\end{equation}
where the entries are linear operators specified as follows.%
\[
A\text{\emph{:}}\mathcal{M}_{1}\rightarrow\mathcal{M}_{1},~B\text{\emph{:}%
}\mathcal{M}_{2}\rightarrow\mathcal{M}_{1}\text{\emph{;}}%
\]
and
\[
C\text{\emph{:}}\mathcal{M}_{1}\rightarrow\mathcal{M}_{2},~D\text{\emph{:}%
}\mathcal{M}_{2}\rightarrow\mathcal{M}_{2}\text{.}%
\]

If $\dim\mathcal{M}_{1}=1$, and $\mathcal{M}_{1}=\mathbb{C}w$ for some
$w\in\mathcal{M}$, then we will identify the operators $\mathcal{M}%
_{1}\rightarrow\mathcal{M}$ with $\mathcal{M}$ via $T_{\eta}$:$\mathbb{C}%
\ni\rightarrow z\eta$, where $\eta\in\mathcal{M}$. the adjoint operator is
$T_{\eta}^{\star}x=\langle\eta|x\rangle w,$~$x\in\mathcal{M}$.

\begin{lemma}
\label{Lem4.1}Let $\mathcal{M}$ be a finite-dimensional complex Hilbert space,
with $d=\dim\mathcal{M}$. Let $F\in L\left(  \mathcal{M}\right)  $, and let
$a\in\mathbb{C}$ satisfy the following four conditions:\medskip

\noindent$\left(  i\right)  $ $a\in\operatorname*{spec}\left(  F\right)  ;$

\noindent$\left(  ii\right)  $ $\left\vert a\right\vert >\max\left\{
\left\vert \lambda\right\vert |\lambda\in\operatorname*{spec}\left(  F\right)
\diagdown\left\{  a\right\}  \right\}  ;$

\noindent$\left(  iii\right)  $ the algebraic multiplicity of $a$ is one$;$

\noindent$\left(  iv\right)  $ there is a $w\in\mathcal{M}$, $\left\Vert
w\right\Vert =1$, such that $F^{\ast}w=\bar{a}w$.\medskip

Then there is a unique $\xi\in\mathcal{M}$ such that
\begin{equation}
\langle w|\xi\rangle=1\text{ and }F\xi=a\xi\text{.} \label{Eq58}%
\end{equation}

Moreover,
\begin{equation}
\lim_{n\rightarrow\infty}a^{-n}F^{n}x=\langle w|x\rangle\xi\text{ for all
}x\in\mathcal{M}\text{.} \label{Eq59}%
\end{equation}

\end{lemma}

\begin{remark}
\label{Rem4.1}There is a constant $C$ independent of $d=\dim\mathcal{M}$ and
$x$, such that
\begin{equation}
\left\Vert a^{-n}F^{n}x-\langle w|x\rangle\xi\right\Vert \leq Cn^{d-1}%
\max\left\{  \left\vert \frac{s}{a}\right\vert ^{n}|s\in\operatorname*{spec}%
\left(  F\right)  \diagdown\left\{  a\right\}  \right\}  \label{Eq60}%
\end{equation}

\end{remark}

\begin{proof}
(Lemma \ref{Lem4.1}.) Set $\mathcal{M}^{\perp}$:$=\mathcal{M}\ominus
\mathbb{C}w=\left\{  x\in\mathcal{M}|\langle w|x\mathcal{\rangle}=0\right\}
$. Then
\begin{equation}
\mathcal{M}=\mathbb{C}w\oplus\mathcal{M}\text{,} \label{Eq61}%
\end{equation}
and we get the resulting block-matrix representation of $F$,
\begin{equation}
F=\left(
\begin{array}
[c]{ll}%
a & 00\cdots0\\
\eta & G
\end{array}
\right)  \label{Eq62}%
\end{equation}
where $a$ is the number in (i), the vector $\eta\in\mathcal{M}^{\perp}$, and
operator $G\in L\left(  \mathcal{M}^{\perp}\right)  $, are uniquely determined.

As a result, we get the factorization
\begin{equation}
\det\left(  \lambda-F\right)  =\left(  \lambda-a\right)  ~\det\left(
\lambda-G\right)  \label{Eq63}%
\end{equation}
for the characteristic polynomial. Assumptions (ii) and (iii) imply
\begin{equation}
\operatorname*{spec}\left(  F\right)  \diagdown\left\{  a\right\}
=\operatorname*{spec}\left(  G\right)  \text{;} \label{Eq64}%
\end{equation}
and in particular, we note that $a$ is not in the spectrum of $G$. Hence the
inverse $\left(  a-G\right)  ^{-1}$ is well defined, and $\left(  a-G\right)
^{-1}\in L\left(  \mathcal{M}^{\perp}\right)  $.

We claim that the vector
\begin{equation}
\xi=w+\left(  a-G\right)  ^{-1}\eta\label{Eq65}%
\end{equation}
satisfies the conditions in (\ref{Eq58}).

First note that $\left(  a-G\right)  ^{-1}\eta\in\mathcal{M}^{\perp}$, so
$\langle w|\xi\rangle=\langle w|w\rangle=\left\Vert w\right\Vert ^{2}=1$. Moreover,

$%
\begin{array}
[c]{l}%
F\xi=aw+\eta+G\left(  a-G\right)  ^{-1}\eta\\
\quad\;\,=aw+a\left(  a-G\right)  ^{-1}\eta\\
\quad\;\,=a\xi\text{,}%
\end{array}
$\newline which proves the second condition in (\ref{Eq58}). Uniqueness of the
vector $\xi$ in (\ref{Eq58}) follows from (\ref{Eq64}).

Using the matrix representation (\ref{Eq62}), we get
\[
F^{2}=\left(
\begin{array}
[c]{ll}%
a^{2} &
\begin{array}
[c]{l}%
00\cdots0\\
G^{n}%
\end{array}
\\
a\eta+G\eta &
\begin{array}
[c]{l}%
00\cdots0\\
G^{n}%
\end{array}
\end{array}
\right)
\]
and by induction,
\begin{align}
F^{2}  &  =\left(
\begin{array}
[c]{lc}%
a^{n} & 00\cdots0\\
a^{n-1}\eta+a^{n-2}G\eta+\cdots+G^{n-1}\eta\, & G^{n}%
\end{array}
\right) \label{Eq66}\\
&  =\left(
\begin{array}
[c]{lc}%
a_{n} & 00\cdots0\\
\left(  a^{n}-G^{n}\right)  \left(  a-G\right)  ^{-1}\eta\qquad\qquad & G^{n}%
\end{array}
\right)  \text{.}\nonumber
\end{align}

Hence, if we show that
\begin{equation}
\lim_{n\rightarrow\infty}a^{-n}G^{n}=0\text{,} \label{Eq67}%
\end{equation}
then the desired conclusion (\ref{Eq59}) will follow. Using the matrix form
(\ref{Eq66}), the conclusion (\ref{Eq59}) reads
\begin{equation}
\lim_{n\rightarrow\infty}a^{-n}F^{n}=\left(
\begin{array}
[c]{lc}%
1 & 00\cdots0\\
\left(  a-G\right)  ^{-1}\eta & 0
\end{array}
\right)  \label{Eq68}%
\end{equation}

In proving (\ref{Eq67}), we will make use of the Jordan-form representation
for $G$. Jordan's theorem applied to $G$ yields three operators $D$, $V$,
$N\in L\left(  \mathcal{M}^{\perp}\right)  $ with the following
properties:\medskip

\noindent(1) $D$ is a diagonal matrix with the numbers $\operatorname*{spec}%
\left(  F\right)  \diagdown\left\{  a\right\}  $ down the diagonals;

\noindent(2) $V$ is invertible;

\noindent(3) $N$ is nilpotent: If $d-1=\dim\left(  \mathcal{M}^{\perp}\right)
$ then $N^{d-1}=0$;

\noindent(4) $\left[  N,D\right]  =ND-DN=0$;

\noindent(5) $G=V\left(  D+N\right)  V^{-1}$.\medskip

Let $x\in\mathcal{M}^{\perp}$, and let $n\geq d$. Using (2)--(5), we get
$a^{-n}G^{n}x=Va^{n}\left(  D+N\right)  ^{n}V^{-1}x=\sum_{i=0}^{d-2}\binom
{n}{i}Va^{-n}D^{n-i}N^{i}v^{-1}x$. But the matrix $a^{-n}D^{n-i}$ is diagonal
with entries $\left\{  a^{-n}s^{n-i}|s\in\operatorname*{spec}\left(  F\right)
\diagdown\left\{  a\right\}  \right\}  ,~0\leq i<d-1$. Using finally
assumption (ii), we conclude that
\begin{equation}
\lim_{n\rightarrow\infty}\binom{n}{i}a^{-n}s^{n-i}=0\text{,} \label{Eq69}%
\end{equation}
and the proof of (\ref{Eq67}) is completed.
\end{proof}

\bigskip

\begin{proof}
(Remark \ref{Rem4.1}) Let the conditions be as stated in the Remark. From the
arguments in the proof of Lemma \ref{Lem4.1}, we see that the two vectors on
the left-hand side in (\ref{Eq60}) may be decomposed as follows:
\begin{equation}
a^{-n}F^{n}x=\langle w|x\rangle w+\left(  1-a^{-n}G^{n}\right)  \left(
a-G\right)  ^{-1}\eta+a^{-n}G^{n}P_{\mathcal{M}^{\perp}}x \label{Eq70}%
\end{equation}
and
\begin{equation}
\langle w|x\rangle\xi=\langle w|x\rangle w+\left(  a-G\right)  ^{-1}%
\eta\text{.} \label{Eq71}%
\end{equation}
Hence, the difference is in $\mathcal{M}^{\perp}$, and
\begin{align*}
\left\Vert a^{-n}F^{n}x-\langle w\left\vert x\right\vert \rangle
\xi\right\Vert  &  =\left\Vert a^{-n}G^{n}\left(  P_{\mathcal{M}^{\perp}%
}x-\left(  a-G\right)  ^{-1}\eta\right)  \right\Vert \\
&  \leq Cn^{d-1}\max\left\{  \left\vert \frac{s}{a}\right\vert ^{n}%
|s\in\operatorname*{spec}\left(  F\right)  \diagdown\left\{  a\right\}
\right\}
\end{align*}
which is the desired conclusion.(\ref{Eq60}).
\end{proof}

\end{document}